\documentclass[trackchanges,twocolumn]{aastex7}

\newcommand{\g}{$\gamma$-ray}
\newcommand{\lat}{\textit{Fermi}-LAT}

\begin{document}

\title{4HWC J2029$+$3641: a Pulsar Wind Nebula Powered by PSR J2030$+$3641?}

\author[sname='OU']{Ziwei Ou}
\affiliation{Tsung-Dao Lee Institute, Shanghai Jiao Tong University, Shanghai 201210, China}
\email[show]{ziwei@sjtu.edu.cn}

\author[sname='Wang']{Jie Wang}
\affiliation{Xinjiang Astronomical Observatory, Chinese Academy of Sciences, Urumqi, 830011, China}
\email[show]{wangjie@xao.ac.cn}

\author[sname='Pei']{Songpeng Pei}
\affiliation{School of Physics and Electrical Engineering, Liupanshui Normal University, Liupanshui 553004, China}
\email{no}

\begin{abstract}

4HWC J2029+3641 is a newly discovered point source detected by HAWC, with no previously identified TeV counterpart. The gamma-ray pulsar PSR J2030+3641, located 0.1$^\circ$ from the source center, is a middle-aged pulsar showing spin parameters similar to Geminga. Using \textit{Fermi}-LAT data spanning from August 2008 to February 2026, we performed binned maximum likelihood spectral analysis in the energy range from 300 MeV to 1 PeV. A phase-resolved analysis was conducted to separate the off-peak and on-peak emissions. No significant spatial extension was found for the off-peak component. The off-peak spectrum exhibits strong curvature and is best described by an exponentially cutoff power-law model. The observed radio-to-gamma phase lag and narrow peak separation favor an outer-gap model for the gamma-ray emission. 

\end{abstract}


\section{Introduction} \label{sec:intro}


The discovery of very-high-energy (VHE) and ultra-high-energy (UHE) \g\ sources has advanced significantly with the operation of ground-based observatories such as the High Altitude Water Cherenkov (HAWC) \citep{Albert2020} observatory and the Large High Altitude Air Shower Observatory (LHAASO) \citep{Cao2024}. These facilities have opened a new window into the non-thermal universe, revealing numerous Galactic sources that emit photons up to PeV energies \citep{Cao2023}. 4HWC J2029$+$3641 (R.A.=307.39$^{\circ}$ Dec.=36.68$^{\circ}$) is a new point source discovery by HAWC \citep{Alfaro2026}. No identified TeV counterpart was found in TeVCat.


A \g\ pulsar J2030$+$3641 locates 0.1$^{\circ}$ away from the center of 4HWC J2029$+$3641 \citep{Smith2023}. With a spin period of 0.2 s, a spin-down luminosity of $3 \times 10^{34}\ \rm erg\ s^{-1}$, and a characteristic age $\tau_c$ of 500 kyr, PSR J2030$+$3641 is a middle-aged neutron star with spin parameters similar to those of the Geminga. The distance is about 1.5 to 3 kpc \citep{Camilo2012}. Given the spatial coincidence between 4HWC J2029$+$3641 and PSR J2030$+$3641, it is natural to investigate whether the TeV emission originates from a pulsar wind nebula (PWN) powered by this pulsar. Middle-aged pulsars are often associated with PWNe, which form when the relativistic wind of particles from the pulsar interacts with the surrounding interstellar medium or the supernova remnant ejecta \citep{Bell1992,Lyubarsky2003}.


PWNe comprise a highly magnetized relativistic plasma, with particles being accelerated at the termination shock where the ram pressure of the pulsar wind is balanced by the pressure of the highly relativistic plasma \citep{Amato2006,Cerutti2020}. These nebulae are efficient particle accelerators, producing synchrotron emission from radio to X-rays and inverse-Compton (IC) scattering of low-energy photons (e.g., cosmic microwave background (CMB), infrared, optical) into the \g\ band \citep{Cao2021b}. Recently, the LHAASO source catalog reported 43 UHE \g\ sources, most of which were identified as PWNe or PWN candidates \citep{Cao2024}. PWN are detected by \lat\ at high-energy \citep{Rousseau2012,Grondin2013} and ground-based telescopes at VHE \citep{Abdalla2018}. Particles from pulsar wind are accelerated at the termination shock, which makes PWN as important \g\ sources. These are generated via IC scattering of high-energy electrons. 

In this work, we analyze \lat\ data (Section~\ref{sec:data}) and divide the pulsar phase into off-pulse and on pulse (Section~\ref{sec:phase}). The off-pulse and on-pulse analysis are presented in Section~\ref{sec:off} and Section~\ref{sec:on}, respectively. We provide discussion in Section~\ref{sec:discus}, including pulsar magnetospheric emission (Section~\ref{subsec:magnetospheric}), leptonic scenario (Section~\ref{subsec:leptonic}), hadronic scenario (Section~\ref{subsec:hadronic}), and possible counterpart of pulsar halo (Section~\ref{subsec:halo}).

\section{Data Analysis}\label{sec:data}

The \lat\ data used in our study span from 2008 August 4 to 2026 February 24. We selected events from the Pass 8 event class and used P8\_R3\_V3 Source instrument response functions (IRFs). All \g\ photons within an energy range of 300 MeV to 1 PeV and within a circular region of interest (ROI) of 15$^{\circ}$ radius centered on PSR J2030+3641 were considered. To minimize \g\ contamination from the Earth’s limb, \g\ photons with a zenith angle $< 95^{\circ}$ were selected.

The spectral results presented in this work were obtained by performing a binned maximum likelihood fit with 40 logarithmically spaced bins in the 0.3 GeV to 1 PeV using the Science Tool \textit{gtlike}. Spectral and spatial models were constructed to include Galactic and isotropic diffuse emission components (“gll\_iem\_v06.fits,”and “iso\_P8R3\_V\_v06.txt,” respectively) as well as known \g\ sources within 15$^{\circ}$ of the pulsar, based on the \lat\ Forth Source Catalog (4FGL-DR4) \citep{Ballet2023}. The spectral parameters of the sources within $4^{\circ}$ of our target were left free. The spectral parameters of other sources included were fixed at the 4FGL-DR4 values. In the pulsar phase-resolved analysis, photons within a specific spin phase interval were selected. 

\section{Phase Selection of Off-peak and On-peak}\label{sec:phase}

\begin{figure*}[ht]
    \centering
    \includegraphics[width=0.75\linewidth]{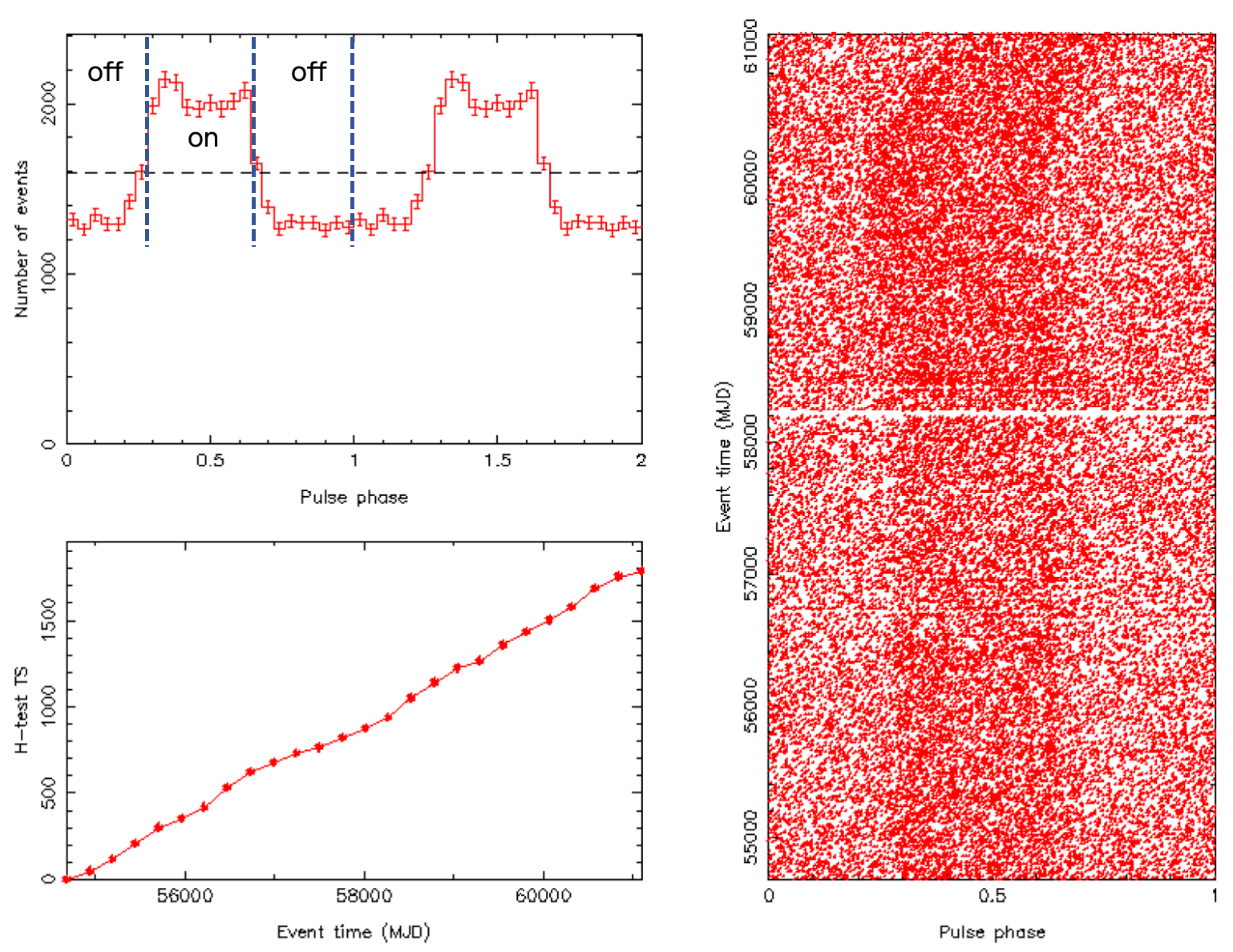}
    \caption{PSR~J2030$+$3641 timing results from \textit{Tempo2} with the \textit{Fermi} plug-in. Top-left panel: phase histogram of the analyzed \lat\ data. Two full rotational phase are shown here. Bottom-left panel: H-test significance (TS) as a function of time. Right panel: pulse phase for each \g\ event vs. time.}
    \label{fig:profile}
\end{figure*}

\begin{figure*}
    \centering
    \includegraphics[width=0.75\linewidth]{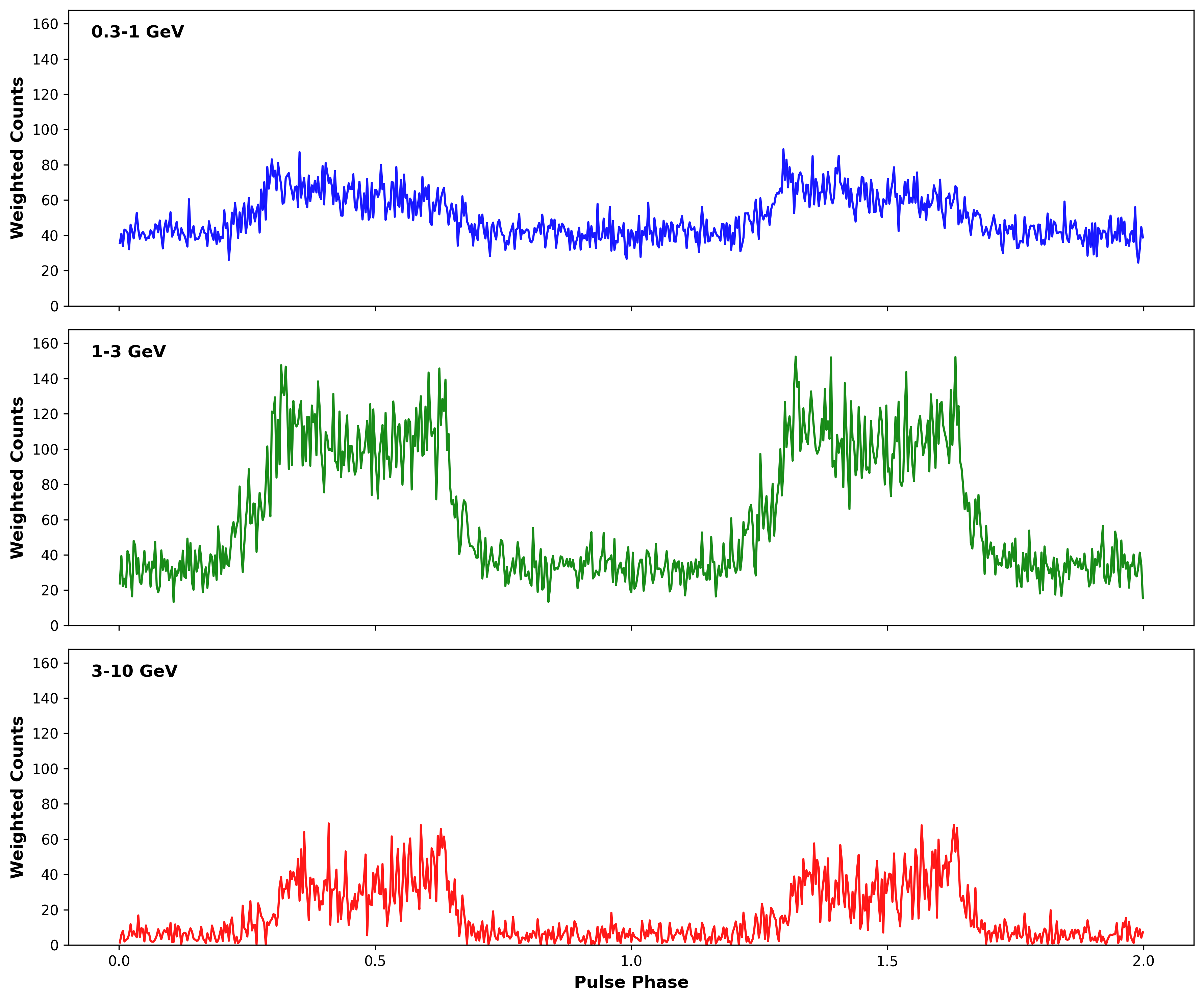}
    \caption{Weighted pulse profile of PSR~J2030$+$3641 at different energies.}
    \label{fig:weight-energy}
\end{figure*}

Photons from PSR J2030$+$3641 within a radius of 0.6$^{\circ}$ and a minimum energy of 300 MeV were selected for maximizing the H-test statistic \citep{deJager2010}. We adopt the ephemeris from 3PC \citep{Smith2023}, which is presented in Table~\ref{tab:ephemeris}. \textit{Tempo2} with \textit{Fermi} plug-in were used to produce \g\ pulse profile \citep{Ray2011}. Thus, we are able to obtain Fourier template profiles for the whole time span, generate the times of arrival (TOAs), and obtain timing solutions by fitting the TOAs with frequency derivatives (Figure~\ref{fig:profile}). The weighted pulse of PSR J2030$+$3641 using \texttt{gtsrcprob} is investigated. Figure~\ref{fig:weight-energy} shows weighted pulse profile of PSR J2030$+$3641 at different energies. These pulsar spin light curves are conducted using a photon weighting technique based on the method of \cite{Kerr2011}. 

\begin{table}[]
    \centering
    \begin{tabular}{ll}
    \hline
    Parameter & Value \\\hline
       Pulsar Name  & PSR~J2030$+$3641 \\
       R.A.  & 20:30:00.265 \\
       Dec.  & 36:41:27.118 \\
       MJD range & 55318-58486 \\
       Pulse frequency $\rm s^{-1}$ & 4.9967688 \\
       First derivative of frequency $\rm s^{-2}$ & $-1.62307\times10^{-15}$\\
       Epoch of frequency (MJD) & 56712 \\
    \hline
    \end{tabular}
    \caption{Timing Ephemeris of PSR~J2030$+$3641}
    \label{tab:ephemeris}
\end{table}

We began by deconstructing the pulsed light curve into simple Bayesian Blocks using the same algorithm described in the 2PC \citep{Abdo2013} and \cite{Caliandro2013}. The off-peak interval in our analysis is then defined to be at $\phi$ = (0-0.26) \& (0.69-1), covering 57\% of the total revolution, which is consistent with the ephemeris of the 3PC. The on-peak phases are thus located at $\phi$ = 0.26-0.69.

\section{Off-peak Analysis}\label{sec:off}

\begin{figure*}
    \centering
    \includegraphics[width=0.48\linewidth]{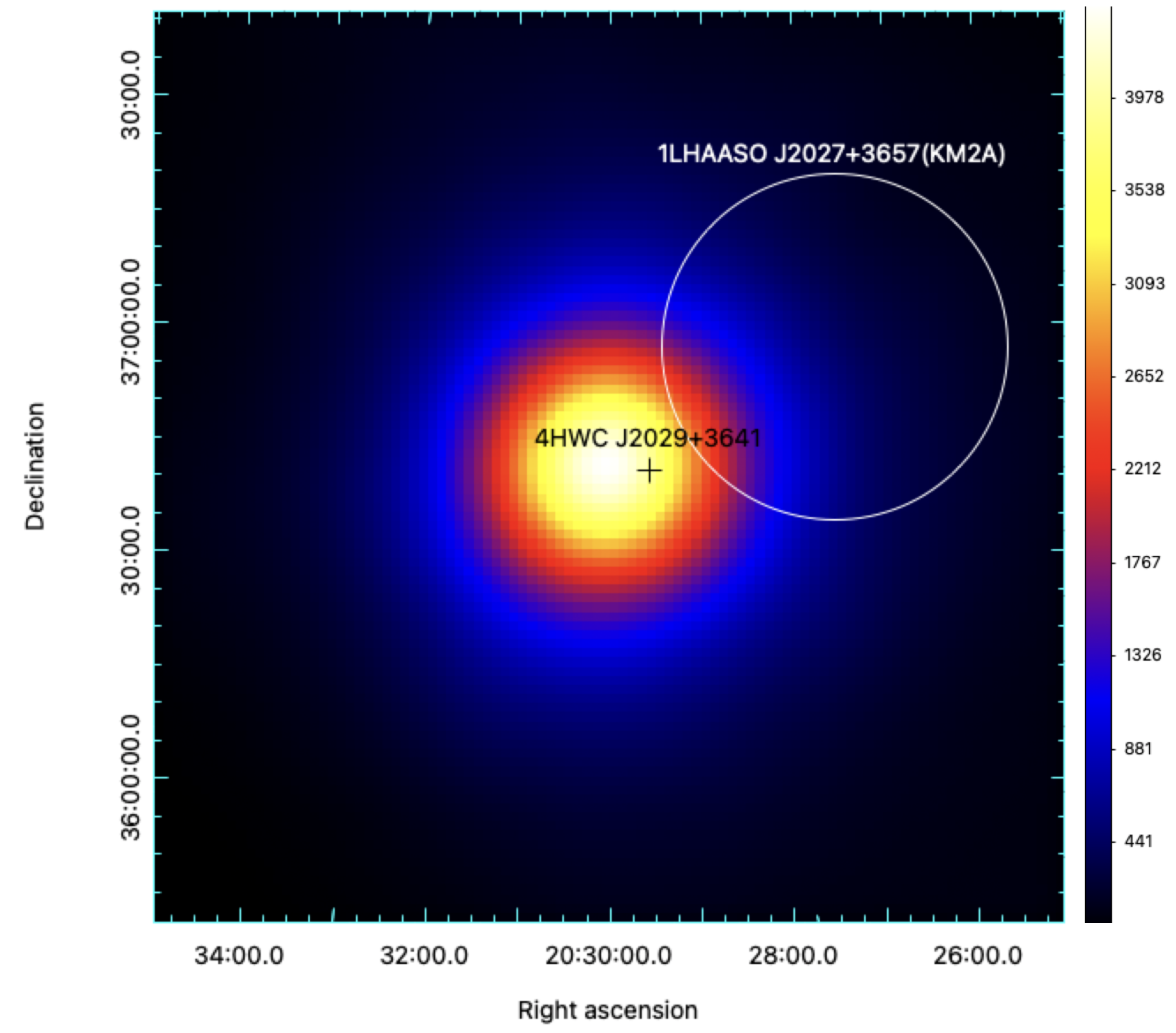}
    \includegraphics[width=0.48\linewidth]{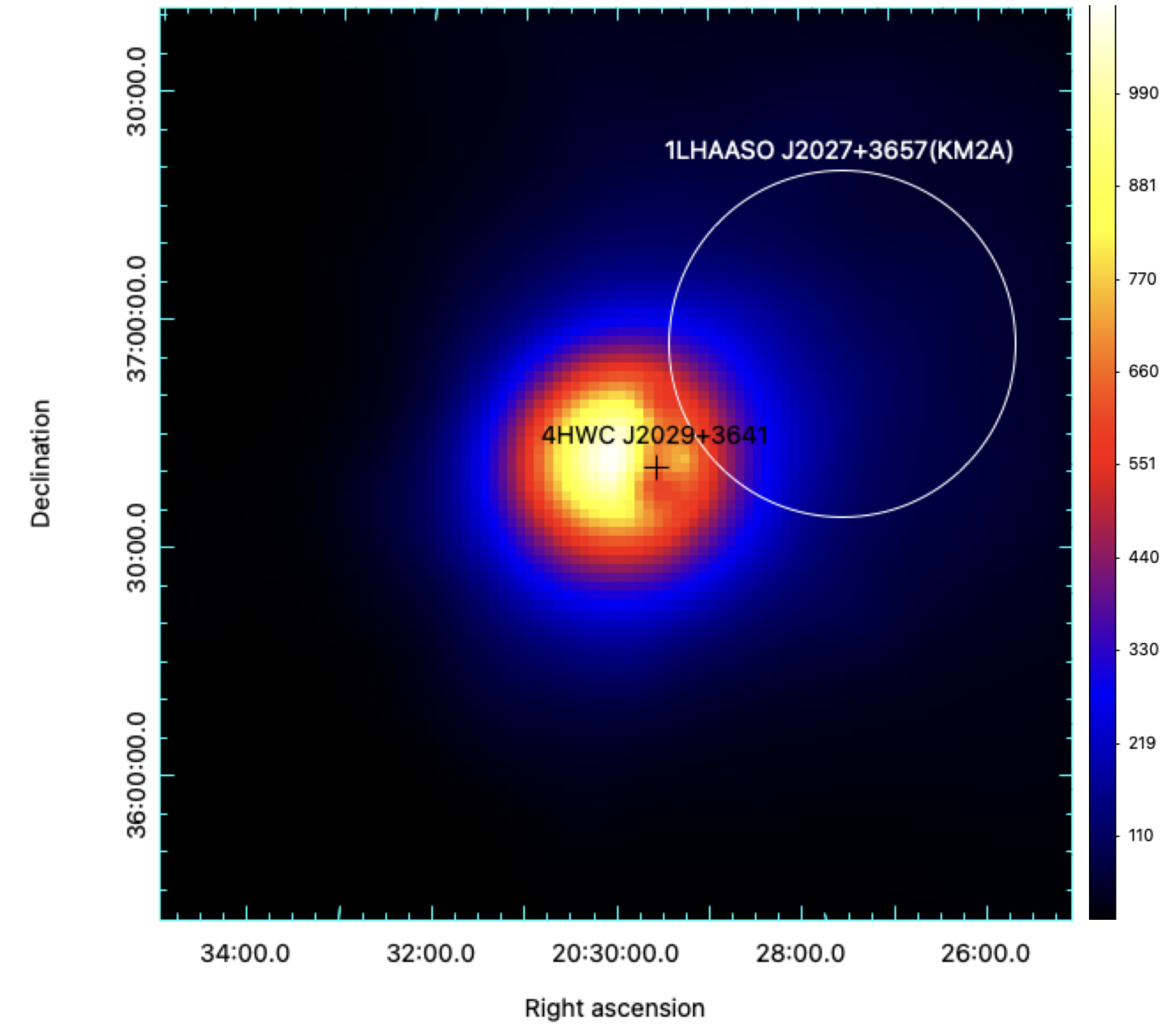}
    \caption{$2^{\circ} \times 2^{\circ}$ TS map (300 MeV to 1 PeV) of whole phase (left) and off-pulse (right) center at the position of PSR~J2030$+$3641. 4HWC~J2029$+$3641 and 1LHAASO~J2027+3657 are presented as black cross and white circle.}
    \label{fig:ts-map}
\end{figure*}

The prefactor parameters of all sources were scaled to the width of the spin phase interval. The $2.0^{\circ} \times 2.0^{\circ}$ test statistic (TS) map were calculated for the source region from the whole LAT data and the off-pulse data, which were presented in the Figure~\ref{fig:ts-map}. The TS is adopted to estimate the significance of \g\ sources, which is defined by TS = 2 (ln $L_1$ - ln $L_0$), where $L_1$ and $L_0$ represent maximum likelihood values for background with target source and without target source. We conduct the analysis by searching radius of PSR~J2030$+$3641 in the off-pulse phase. A source extension analysis is executed by performing a likelihood ratio test with respect to the point source and the extended source: TS$_{\rm ext}$ = -2 $\left( \mathrm{ln}\left(L_{\rm PS}\right) - \mathrm{ln}\left(L_{\rm ext}\right) \right)$. A radial Gaussian model is used to obtain TS$_{\rm ext}$. The best-fit extension values are determined by performing a likelihood profile scan over the 68\% containment and fitting for the extension which maximizes the model. We got TS$_{\rm ext}$ = 2.1 for off-pulse and upper limit on the radius of 0.08$^{\circ}$.

For the spectrum of off-pulse emission, we consider a simple power-law (PL) and a an exponential cutoff power-law model (PLSC). PL model is defined by: 

\begin{equation}
    \frac{dN}{dE} = N_0 (\frac{E}{E_0})^{-\Gamma}
\end{equation}

where $N_0$, $E_0$ and $\Gamma$ are normalization, pivot energy and photon index, respectively. PLSC model is defined by: 

\begin{equation} \label{eq:plsc}
    \frac{dN}{dE} = N_0 \left( \frac{E}{E_0} \right)^{-\Gamma-\frac{d}{2}\ln\frac{E}{E_0} - \frac{db}{6} \ln^2\frac{E}{E_0} - \frac{db^2}{24} \ln^3\frac{E}{E_0}}
\end{equation}

where $\Gamma$ $d$ and $b$ are the photon index, the local curvature at $E_0$, and the parameter described shape of the exponential cutoff, respectively. We fixed $b=2/3$ as provided by \lat\ pulsar catalogs. We obtain a TS value of 210 for off-pulse (Table~\ref{tab:fit-result}). 

The spectral shape helps us determine the nature of \g\ sources. To validate the spectral model, we test for spectral curvature, which reveals deviations from a PL spectrum for each source using a likelihood ratio test. For a PLSC model, it gives: TS$_{\rm PLSC}$ = -2 $\left( \mathrm{ln}\left(L_{\rm PL}\right) - \mathrm{ln}\left(L_{\rm PLSC}\right) \right)$. Once TS$_{\rm PLSC}$ is larger than 25, we perform a spectral fit using the PLSC model, otherwise, the PL model is adopted. The curvature test yields a significant value TS$_{\rm PLSC} = 247$ for off-pulse, indicating that the PLSC model should be used to describe off-pulse. The best-fit parameters are listed in Table~\ref{tab:fit-result}. Corresponding spectra can be found in Figure~\ref{fig:spectrum} (left). 

\begin{table*}[]
    \centering
    \begin{tabular}{ccccccc}
    \hline
        & Phase Range & Flux & $\Gamma$ & $d$ & TS & log(L) \\
        & & ($\rm ph\ s^{-1}\ cm^{-2}$) & & & & \\\hline
        Phase Averaged & -- & $(2.06 \pm 0.07) \times 10^{-08}$ & $1.95 \pm 0.03$ & $0.83 \pm 0.05$ & 5079 & -9688355.462\\\hline
        on-pulse & 0.26-0.69 & $(9.72 \pm 0.17) \times 10^{-09}$ & $1.81 \pm 0.04$ & $0.81 \pm 0.05$ & 1403 & -154129.9931\\\hline
        off-pulse & 0-0.26 \& 0.69-1 & $(1.15 \pm 0.12) \times 10^{-09}$ & $2.71 \pm 0.09$ & - & 210 & -269178.3188\\
    \hline
    \end{tabular}
    \caption{Binned likelihood analysis results for whole data, on-pulse and off-pulse of PSR~J2030$+$3641. These include flux, photon index $\Gamma$, distance, TS values, and log-likelihood values.}
    \label{tab:fit-result}
\end{table*}

\section{On-peak Analysis}\label{sec:on}

\begin{figure*}
    \centering
    \includegraphics[width=0.48\linewidth]{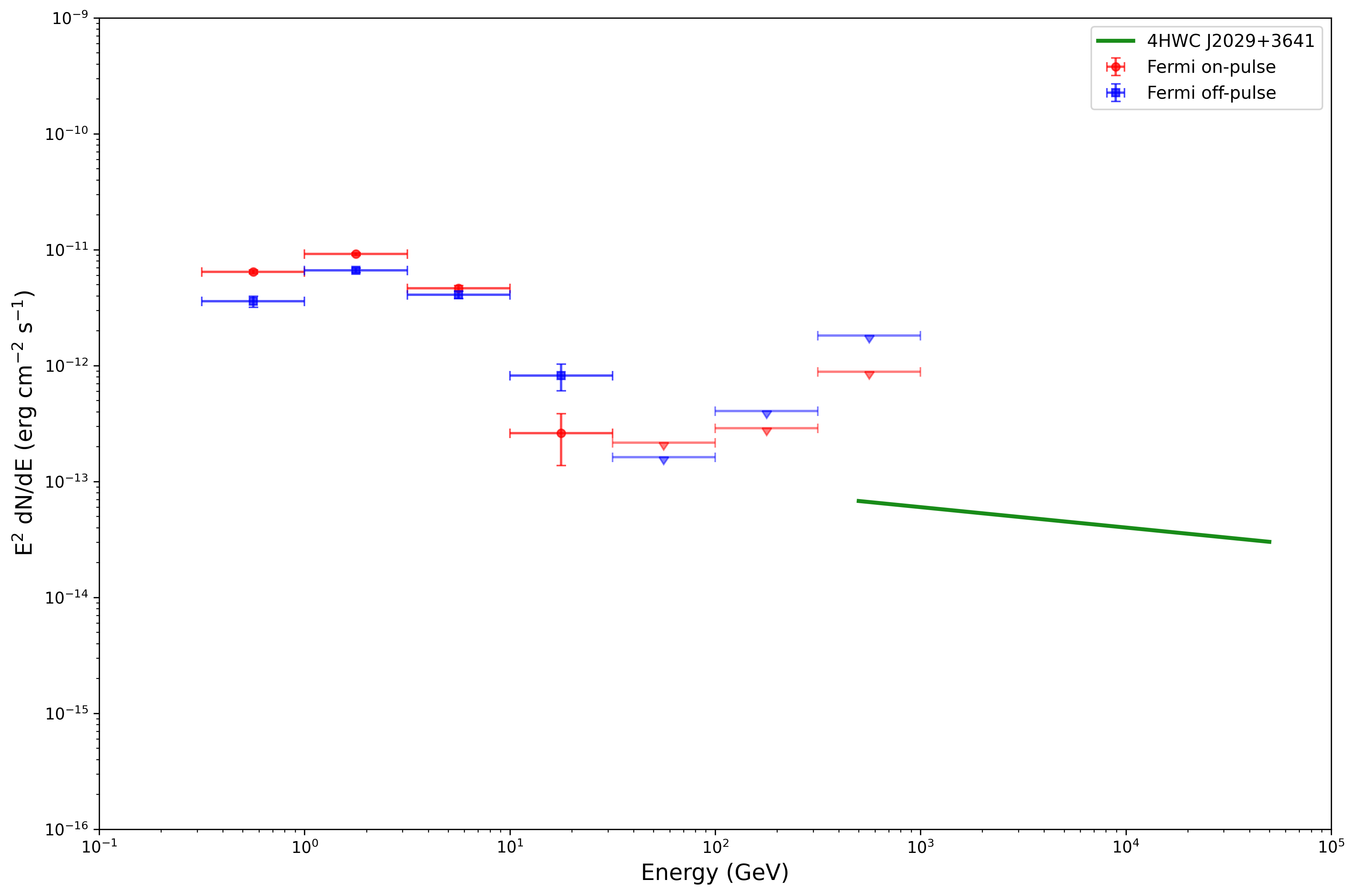}
    \includegraphics[width=0.48\linewidth]{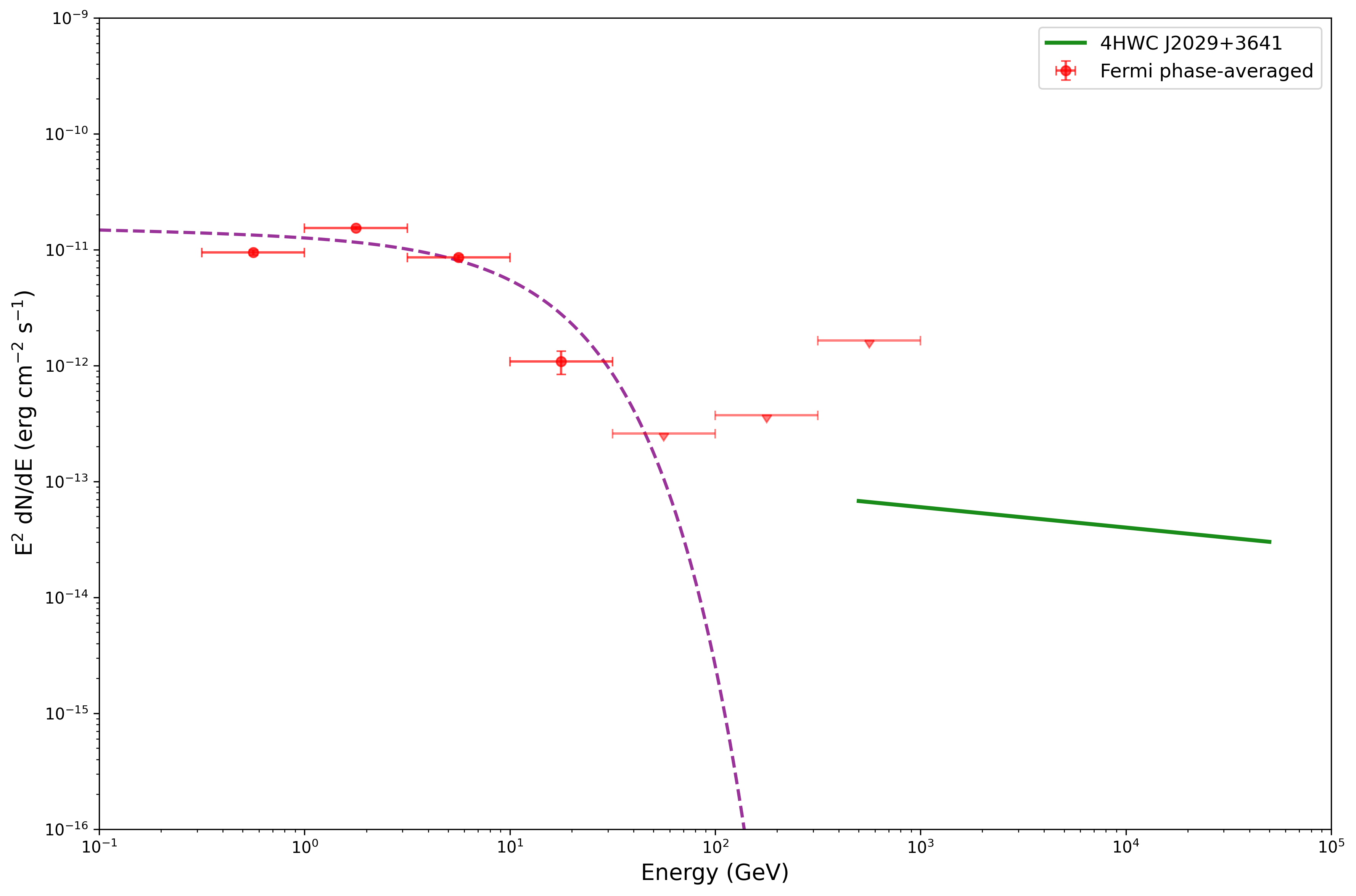}
    \caption{Left: \lat\ spectra of PSR~J2030$+$3641 during on-peak and off peak. Right: phase-averaged \lat\ spectrum of PSR~J2030$+$3641.}
    \label{fig:spectrum}
\end{figure*}

We have considered a PLSC model for modeling the on-peak emission and phase-averaged emission of PSR~J2030$+$3641. The phase-averaged \lat\ spectrum can be found in Figure~\ref{fig:spectrum} (right). The best-fit parameters are shown in Table~\ref{tab:fit-result}. Adopting the best-fit spectral model derived, we calculated the probabilities for photons to come from PSR~J2030$+$3641 within a radius of 1$^{\circ}$. To locate the peak, we fitted the light curve with two asymmetric Lorentzian functions plus a constant. The fitted constant accounts for the background.

\section{Discussions} 
\label{sec:discus}

\subsection{The Magnetospheric Emission from PSR J2030$+$3641} 
\label{subsec:magnetospheric}

The derived \g\ conversion efficiency $\eta$ for PSR J2030$+$3641 is about $142 \pm 6 \pm 100\%$. This seemingly unphysical result is most likely driven by the systematic uncertainty in the pulsar distance as discussed in \cite{Camilo2012}, because $L_{\gamma}$ scales as the square of the distance and the nominal value of $\dot{E}$ itself depends on distance. The second, much larger error bar is deliberately included to accommodate such systematic effects, notably the uncertainty in the distance (DM‑based distance $\sim 460$ pc versus dust‑scattering halo distance $\sim 400$) and possibly also the beaming factor. The lower part of the allowed range already places this pulsar among the most efficient \g\ emitters, comparable to the “Geminga‑like” pulsars. Thus, the measurement indicates that PSR J2030+3641 converts a very large fraction of its spin‑down power into high‑energy radiation, highlighting the need for an accurate geometric distance to pin down the true conversion efficiency.

The observed \g\ peak separation of $\Delta = 0.32$ in phase is relatively narrow compared to the canonical value of $\sim$0.4-0.5 seen in many young \g\ pulsars \citep{Abdo2013}. This indicates a viewing geometry where the line of sight cuts the emission cone close to its center \citep{Watters2009}. Modeling of the light curve and polarization profile for PSR J2030$+$3641 has constrained the magnetic inclination angle $\alpha$ to the range of about 38$^{\circ}$ to 54$^{\circ}$, implying a nearly orthogonal rotator ($\alpha \sim 50^{\circ}$) \citep{Camilo2012}. Such a geometry is crucial for determining the three‑dimensional structure of the magnetosphere and the location of particle acceleration regions.

The measured phase lag of radio/$\gamma$ phase gives $\delta \approx 0.3$. The positive lag (\g\ arriving later) favors an outer‑magnetospheric origin for the high‑energy emission, with the \g\ production site located at a larger altitude than the radio emission region. This phase lag may translate into an emission height for the radio pulses. The consistency of this lag with predictions of the outer gap (OG) model further supports the scenario where the \g\ are produced in the outer magnetosphere via curvature radiation from relativistic particles accelerated in a vacuum gap \citep{Venter2012,Kalapotharakos2014}.

The combination of a moderate peak separation and a clear phase lag provides a discriminating test between competing emission models \citep{Pierbattista2016}. While the OG model naturally accommodates both the narrow peak separation and the positive lag, the two‑pole caustic (TPC) model \citep{Muslimov2004} would require extreme line‑of‑sight cuts to reproduce $\Delta$ and generally struggles to explain the observed lag magnitude. Therefore, the phase parameters of PSR J2030$+$3641 strongly favour the OG picture of high‑energy emission from the outer magnetosphere, with the radio emission arising from closer to the stellar surface. These results highlight the power of combined phase‑resolved \g\ and radio observations in mapping the three‑dimensional emission geometry and testing magnetospheric models.

\subsection{Leptonic Scenario}
\label{subsec:leptonic}

PWNe have been recognized as efficient electron factories. They are powered by energetic pulsars, which inject relativistic electrons and positrons in their magnetosphere. By considering the leptonic scenario, we include synchrotron radiation, IC (with the CMB as well as with IR/optical photon fields), self-synchrotron Compton, and bremsstrahlung to compute the radiation these processes.

\cite{Wilhelmi2022} explores the potential of young, energetic pulsars to power UHE sources particularly those recently discovered by LHAASO with spectra extending beyond 100 TeV \citep{Cao2021a}. The absolute maximum energy the particles depend on the maximum potential drop of pulsar. The key factor determining whether a pulsar can accelerate electrons to PeV energies is its spin-down luminosity $\dot{E}$. The authors derive an absolute maximum photon energy, $E_{\gamma, \rm max} \approx 0.9 \dot{E}_{36}^{0.65}\ \rm PeV$, based on the maximum potential drop available in the pulsar magnetosphere under ideal MHD flow \citep{Olmi2019}. For most pulsars, this potential-drop limit is more restrictive than synchrotron losses, meaning the pulsar’s rotational power directly sets the upper bound on achievable particle energies. Considering $\eta_{\rm b} = \eta_{e} \eta_{\rm B}^{1/2}$ which is defined by \cite{Wilhelmi2022}, we obtain $\eta_{\rm b} = 1.91$. Assuming $\eta_{e} = \eta_{\rm B} = 1$, we have $E_{\gamma,\max}=99.7\ \rm TeV$, corresponding to $E_{e,\max}=363\ \rm TeV$ (Table~\ref{tab:max-energy}).

\begin{table}[]
    \centering
    \begin{tabular}{c|c|c|c}
    \hline
HAWC Source & Pulsar & $E_{\gamma,\rm max}$ & $E_{e,\rm max}$\\
& & (PeV) & (PeV) \\ \hline
J0538+2804	& J0538+2817	& 0.13 	& 0.44 \\
J0542+2315	& B0540+23	    & 0.11 	& 0.40 \\
J0615+2214	& B0611+22	    & 0.15 	& 0.50 \\
J0622+3759	& J0622+3749	& 0.09 	& 0.33 \\
J0631+1036	& J0631+1036	& 0.28 	& 0.82 \\
J0633+1719	& J0633+1746	& 0.10 	& 0.36 \\
J0634+0628	& J0633+0632	& 0.23 	& 0.69 \\
J0701+1412	& J0659+1414	& 0.11 	& 0.39 \\
J1740+0950	& J1740+1000	& 0.35 	& 0.96 \\
J1804-2133	& B1800-21	    & 1.51 	& 2.98 \\
J1809-1923	& J1809-1917	& 1.31 	& 2.67 \\
J1813-1244	& J1813-1246	& 2.96 	& 5.00 \\
J1834-0828	& B1830-08	    & 0.63 	& 1.53 \\
J1857+0200	& B1855+02	    & 0.08 	& 0.30 \\
J1912+1013	& J1913+1011	& 1.79 	& 3.39 \\
J1914+1151	& J1915+1150	& 0.60 	& 1.47 \\
J1923+1631	& J1924+1639	& 0.08 	& 0.32 \\
J1928+1843	& J1930+1852	& 4.43 	& 6.81 \\
J1930+1852	& J1930+1852	& 4.43 	& 6.81 \\
J1932+1917	& J1932+1916	& 0.50 	& 1.28 \\
J2005+3056	& J2006+3102	& 0.34 	& 0.95 \\
J2018+3641	& J2017+3625	& 0.05 	& 0.22 \\
J2021+4036	& J2021+4026	& 0.22 	& 0.68 \\
J2026+3327	& J2028+3332	& 0.10 	& 0.37 \\
J2029+3641	& J2030+3641	& 0.10 	& 0.36 \\
\hline
    \end{tabular}
    \caption{HAWC sources (4HWC) and associated pulsars, with maximum \g\ energy and maximum electron energy.}
    \label{tab:max-energy}
\end{table}

To serve as a UHE \g\ source, a pulsar should satisfy several conditions. First, the inferred electron energy from LHAASO observations should not exceed the theoretical maximum set by $\dot{E}$. Second, the nebular magnetic field should be low enough (typically a few tens of $\mu$G) to avoid severe synchrotron losses and to allow sufficient conversion of spin-down power into \g\ emitting electrons via IC scattering off the CMB. Thus, bright young pulsars with high $\dot{E}$ and moderate magnetic fields are promising PeVatrons.

\subsection{Hadronic Scenario}
\label{subsec:hadronic}

The acceleration of protons in PWN and the subsequent production of hadronic \g\ signals are theoretically feasible \citep{Lemoine2015,Mitchell2026}. Hadrons can originate from three main scenarios: (1) ions stripped from the pulsar surface and accelerated in the magnetosphere, particularly if the pulsar was born with a millisecond period; (2) mass-loading of the pulsar wind by ions from the surrounding interstellar medium, which are then accelerated in the wind or at the termination shock \citep{Lyutikov2003,Morlino2015}; and (3) shock mixing or re-acceleration of pre-existing material from the supernova remnant or ambient medium, for example through reverse shock interactions or diffusion back into the PWN \citep{Blondin2001,Ohira2018}. In these scenarios, accelerated protons can interact with ambient gas via pp collisions, producing neutral pions that decay into high-energy \g\ , as well as charged pions that ultimately yield neutrinos. Thus, from a theoretical perspective, PWNe are viable sites for hadronic particle acceleration and the emission of hadronic signals.

However, despite this theoretical viability, current observational evidence suggests that hadronic emission is at most sub-dominant in most PWNe. In addition, stacking analyses by the IceCube collaboration have found no significant neutrino excess from a sample of TeV-bright PWNe, placing upper limits on the hadronic fraction \citep{Aartsen2020}. For individual sources like the Crab Nebula, lepto-hadronic models remain speculative \citep{Cao2021b,Aharonian2024}, and constraints from both \g\ spectral features (e.g., the absence of a clear pion-decay bump) and neutrino non-detections indicate that any hadronic component must be energetically sub-dominant. Therefore, while hadronic acceleration in PWNe is not ruled out and remains an exciting possibility \citep{Cao2021b,Cao2024}.

\subsection{Pulsar Halo}
\label{subsec:halo}

\cite{Giacinti2020} investigates the conditions under which TeV \g\ halos form around pulsars, distinguishing them from PWN. A key factor is the energy density of relativistic electrons $\epsilon_e$ relative to that of the interstellar medium $\epsilon_{\rm ISM}$. Using two estimators—one based on pulsar spin-down properties and the other on TeV \g\ luminosity—the authors find that most TeV-bright sources have $\epsilon_e \gtrsim \epsilon_{\rm ISM}$ indicating that their emission originates from within the PWN. Only when $\epsilon_e \lesssim \epsilon_{\rm ISM}$ (e.g., for Geminga and PSR B0656$+$14 \citep{Abeysekara2017,DiMauro2019}) can the emission be classified as a true halo, where electrons have escaped into the unperturbed ISM. Considering $\epsilon_{\rm e} = E_{\rm inj}/V$ and $E_{\rm inj} = \dot{E}/\tau_{\rm c}$, we obtain $R > 27.1\ \rm pc$ for PSR J2030$+$3641. This is corresponding to 0.22$^{\circ}$ as the distance of pulsar is 6.95 kpc.

The evolutionary stage of the pulsar system also plays a critical role in halo formation. Halos are expected only at late times ($t \gtrsim 100$ kyr), after the pulsar has escaped its SNR and electrons diffuse into the ISM \citep{Hooper2018}. PSR J2030$+$3641 satisfies such age condition of being halo. In contrast, younger systems still confine most high-energy electrons within the PWN or surrounding SNR. Although old, low-power pulsars are numerous, their lower surface brightness and rapid cooling of high-energy electrons mean that halos contribute little to the total TeV \g\ luminosity from pulsars.

\section{Conclusions}
\label{sec:conclu}

Based on the \lat\ data, we have performed a detailed phase-resolved spectral analysis of the region around 4HWC J2029+3641 and its spatial coincidence with the \g\ pulsar PSR J2030+3641. Our main findings are as follows. The off-pulse emission (which isolates the putative pulsar wind nebula from the magnetospheric contribution) shows no significant spatial extension, with TS$_{\rm ext}$ = 2.1. This yields an upper limit on the source radius of $< 0.1^{\circ}$, consistent with a compact PWN rather than a diffuse halo. The off-pulse spectrum exhibits strong curvature and is best described by a PLSC model. The on-pulse spectrum is harder.

The curved off-pulse spectrum can be adequately reproduced by IC scattering of relativistic electrons off the CMB and infrared/optical photon fields, with a nebular magnetic field of a few $\mu$G. However, a hadronic scenario can not be ruled out. PSR J2030+3641 converts a very large fraction of its spin-down power into gamma rays, with an apparent efficiency $\eta > 100\%$ when using the nominal DM distance. This paradox likely originates from systematic distance uncertainties and/or a small beaming correction. The observed radio-to-gamma phase lag $\delta \approx 0.3$ and narrow peak separation ($\Delta = 0.32$) favour an outer-gap model for the high-energy emission, placing the \g\ production region near the light cylinder.

\begin{acknowledgments}

Scientific results from data presented in this publication are obtained from HEASARC. Ziwei Ou is supported by the National Natural Science Foundation of China (NSFC, No. 12393853). Jie Wang is supported by the National Natural Science Foundation of China (No. 12573113); the Tianshan Talent Training Program (No. 2023TSYCCX0112); the Tianshan Innovation Team Plan of Xinjiang Uygur Autonomous Region (No. 2025D14014). Songpeng Pei is supported by the  Science and Technology Foundation of Guizhou Province (QKHJCMS[2026]752). 

\end{acknowledgments}

\bibliography{sample7}{}
\bibliographystyle{aasjournalv7}

\end{document}